\documentclass[a4paper,fleqn]{cas-dc}

\usepackage[numbers]{natbib}
\usepackage{color}

\def\12C{{^{12}\mathrm{C}}}
\def\Amax{A_\mathrm{max}}
\def\Ecm{E_{\mathrm{cm}}}
\ifdc
\def\Width{\hsize}
\else
\def\Width{0.5\hsize}
\fi

\begin{document}

\let\WriteBookmarks\relax
\def\floatpagepagefraction{1}
\def\textpagefraction{.001}
\shorttitle{Search for the 6$\alpha$ condensed state in $^{24}$Mg using the $\12C+\12C$ scattering}
\shortauthors{Y. Fujikawa et~al.}


\title [mode = title]{Search for the 6$\alpha$ condensed state in $^{24}$Mg using the $\12C+\12C$ scattering}

\author[kyoto]{Y.~Fujikawa}[orcid=0000-0002-9972-0248]
\cormark[1]
\cortext[cor1]{Corresponding author}
\ead{fujikawa@nh.scphys.kyoto-u.ac.jp}
\author[osaka]{T.~Kawabata}[orcid=0000-0002-7014-2421]
\ead{kawabata@phys.sci.osaka-u.ac.jp}
\author[CYRIC]{S.~Adachi}
\author[kyoto]{S.~Enyo}
\author[osaka]{T.~Furuno}
\author[kyoto,RIKEN]{Y.~Hijikata}
\author[osaka]{K.~Himi}
\author[JAEA]{K.~Hirose}
\author[osaka]{Y.~Honda}
\author[kyoto]{K.~Inaba}
\author[JAEA]{H.~Makii}
\author[RCNP]{K.~Miyamoto}
\author[RCNP]{M.~Murata}
\author[JAEA]{K.~Nishio}
\author[kyoto]{S.~Okamoto}
\author[JAEA]{R.~Orlandi}
\author[osaka]{K.~Sakanashi}
\author[JAEA]{F.~Suzaki}
\author[osaka]{S.~Tsuji}
\author[kyoto]{K.~Yahiro}
\author[kyoto]{J.~Zenihiro}

\address[kyoto]{Department of Physics, Kyoto University, Kitashirakawa-Oiwake, Sakyo, Kyoto 606-8502, Japan}
\address[osaka]{Department of Physics, Osaka University, 1-1 Machikaneyama, Toyonaka, Osaka 560-0043, Japan}
\address[CYRIC]{Cyclotron and Radioisotope Center (CYRIC), Tohoku University, 6-3 Aoba, Aramaki, Aoba, Sendai, Miyagi 980-8578, Japan}
\address[RIKEN]{Nishina Center for Accelerator Based Science, RIKEN, 2-1 Hirosawa, Wako, Saitama 351-0198, Japan}
\address[JAEA]{Advanced Science Research Center, Japan Atomic Energy Agency (JAEA), 2-4 Shirakata, Tokai, Ibaraki 319-1195, Japan}
\address[RCNP]{Research Center for Nuclear Physics (RCNP), Osaka University, 10-1 Mihogaoka, Ibaraki, Osaka 567-0047, Japan}

\begin{abstract}
  We searched for the 6$\alpha$-condensed state in $^{24}$Mg by measuring the $\12C+\12C$ scattering with the SAKRA Si detector array at $\Ecm=17.5$--25.0~MeV.
  By using the invariant-mass method for the detected 3$\alpha$ particles,
  the inclusive cross sections for the $\12C+\12C\to\12C(0^+_2)+X$ and $\12C(3^-_1)+X$ reactions were determined.
  In addition, the missing-mass spectroscopy was successfully utilized to determine the excitation energy of the residual $\12C$ nucleus
  and the exclusive cross sections for the $\12C+\12C\to\12C(0^+_2)+\12C(0^+_1)$, $\12C(0^+_2)+\12C(2^+_1)$, and $\12C(0^+_2)+\12C(0^+_2)$ reactions.
  In both the inclusive $\12C(0^+_2)+X$ channel and the exclusive $\12C(0^+_2)+\12C(0^+_1)$ channel,
  the cross section peaked at $\Ecm=19.4$~MeV, which correspond to the excitation energy of $E_x=33.3$~MeV in $^{24}$Mg.
  This 19.4-MeV state is a candidate for the 6$\alpha$-condensed state
  because of the agreement of the excitation energy with the theoretical value and its decay property.
  In the exclusive $\12C(0^+_2)+\12C(0^+_2)$ channel, a broad state was observed at $\Ecm=22.5$~MeV,
  which correspond to the excitation energy of $E_x=36.4$~MeV in $^{24}$Mg.
  From the angular distribution of the differential cross section,
  the spin and parity of this 22.5-MeV state was assigned to be $4^+$.
  In addition, a $2^+$ state was suggested at the low-energy side of the 22.5-MeV state.
  Because their excitation energies are higher than the theoretical value of the 6$\alpha$-condensed state,
  these states might be excited states of the 6$\alpha$-condensed state
  such as the $2^+_2$ and $4^+_1$ states in $\12C$.
\end{abstract}


\maketitle

Particle clustering is an important phenomenon in atomic nuclei.
In particular, the alpha cluster, comprising two protons and two neutrons, is the most essential cluster component due to its strongly bound nature.
The most famous alpha-cluster state is the $0^+_2$ state at $E_x=7.65$~MeV in $\12C$, which is called the Hoyle state \cite{Hoyle1953}.
This state is proposed to be an $\alpha$-condensed state in which all the alpha clusters are condensed into the lowest 0$s$ orbit \cite{Tohsaki2001}.
The $\alpha$-condensed state has a dilute structure with a larger radius than the ground state by a factor of about 1.5.
Due to its dilute nature, the momentum distribution of the alpha clusters is concentrated at $k<1~\mathrm{fm}^{-1}$,
resulting in an exceptionally narrow distribution \cite{Yamada2005}.
The $\alpha$-condensed state can be regarded as the ground state of relative motion between alpha clusters, which are dilutely distributed in space.
The $2^+_2$ state at $E_x=9.87$~MeV \cite{Itoh2011,Zimmerman2013} and the $4^+_1$ state at $E_x=13.3$~MeV \cite{Freer2011} in $\12C$ are corresponding to
the Hoyle analogue states where the relative motion between alpha clusters in the 3$\alpha$-condensed state is excited \cite{Funaki2015}.
The $\alpha$-condensed states with dilute density are predicted to exist even in heavier self-conjugate $A=4k$ nuclei up to $k=10$ \cite{Yamada2004}.
It is theoretically expected that the $k\alpha$-condensed states are located above the $k\alpha$-decay thresholds, and
alpha clusters are confined in a shallow potential pocket formed by the interplay of short-range weak nuclear attractive interaction
and long range repulsive Coulomb interaction.
Exploring the existence of the $\alpha$-condensed states in various nuclei is an intriguing endeavor
to verify the theoretical predictions and to establish universal presence of dilute $\alpha$-cluster states \cite{vonOertzen2010,Yamada2012_book}.
The ubiquity of the $\alpha$-condensed states will suggest that the $\alpha$ condensation manifests in the dilute nuclear matter.
The $\alpha$ condensation affects physical properties of the dilute nuclear matter \cite{Satarov2021,Ebran2020}.

We have currently achieved a certain consensus regarding the existence of the $\alpha$-condensed states in $^{8}$Be and $\12C$.
The ground state of $^8$Be and the $0^+_2$ state in $\12C$ are considered to be the 2$\alpha$- and 3$\alpha$-condensed states.
These states are located slightly above the 2$\alpha$- and 3$\alpha$-decay thresholds,
and are well described by the microscopic alpha-cluster models \cite{Funaki2002,Kamimura1981,Uegaki1979}.
The 4$\alpha$-condensed state in $^{16}$O is theoretically predicted to be the $0^+_6$ state \cite{Funaki2008,Yamada2012,Funaki2018},
and the known $0^+$ state at $E_x=15.097\pm0.005$~MeV \cite{Tilley1993} is a candidate for the corresponding state.
Recently, it was reported that this state decays with the almost same probability into the two $^{8}$Be ground states or the $\alpha+\12C(0^+_2)$ state \cite{Barbui2018}.
The $\alpha$-condensed states are expected to decay into the $\alpha$-condensed state in lighter nuclei
because the overlap between the wave functions of the $\alpha$-condensed states in different nuclei should be large.
Thus, this state in $^{16}$O is a strong candidate for the 4$\alpha$-condensed state.

For $^{20}$Ne, several candidates for the 5$\alpha$-condensed state were experimentally proposed.
In Ref.~\cite{Adachi2021}, it was found that
the three states at $E_x=23.6$, 21.8, and 21.2~MeV in $^{20}$Ne are strongly coupled to the candidate for the 4$\alpha$-condensed state at $E_x=15.097\pm0.005$~MeV in $^{16}$O.
This strong coupling between these observed states and the 4$\alpha$-condensed state is the compelling evidence that they are the candidates for the 5$\alpha$-condensed state.
However, the spins and parities of these candidate states were not determined in Ref.~\cite{Adachi2021}
although they could be another strong evidence of the 5$\alpha$-condensed state.
The spin and parity of the $\alpha$-condensed state must be $0^+$ because all the alpha clusters are condensed into the 0$s$ orbit.
The other candidate for the 5$\alpha$-condensed state was proposed to be the $0^+$ state at $E_x=22.5$~MeV in $^{20}$Ne \cite{Swartz2015}.
This state is not described by the shell model, and its excitation energy is close to the theoretical value of $E_x=21.14$~MeV \cite{Yamada2004}.
However, because of the high level density around the expected excitation energy, it is difficult to conclude this state
to be the 5$\alpha$-condensed state from the excitation energy only.
In either case, theoretical efforts to interpret the experimental data are desired to pin down the 5$\alpha$-condensed state.

For $^{24}$Mg, no theoretical and experimental candidate for the 6$\alpha$-condensed state has been proposed.
In Ref.~\cite{Kawabata2013}, inelastic alpha scattering was employed to search for the 6$\alpha$-condensed state.
Several bump structures were observed in the excitation-energy spectrum of $^{24}$Mg for coincidence events where $^{8}$Be, as the 2$\alpha$-condensed state,
was detected at the same time with an inelastically scattered alpha particle.
However, it was statistically too poor to propose candidates for the 6$\alpha$-condensed state.
Thus, further measurements to search for the 6$\alpha$-condensed state are still needed.
One of the difficulties in searching for the $\alpha$-condensed states by measuring decay particles
is angular coverage of decay-particle detectors.
In the case of the inelastic alpha scattering under normal kinematic conditions,
decay particles are emitted over all solid angles in the laboratory frame because the target nuclei are hardly recoiled.
Therefore, the decay-particle detectors are required to cover large solid angle around the target.
On the other hand, heavy-ion beams are useful for measuring decay particles because decay particles are boosted and focused on forward angles owing to the center-of-mass motion.
Actually, decay-particle measurements in heavy-ion induced reactions were utilized to search for $\alpha$-cluster states
\cite{Barbui2018,Wuosmaa1993,Wuosmaa1994,Freer1997,Vadas2015,Bishop2019,Kolata1980}.
In Ref.~\cite{Bishop2019}, the $\12C({^{16}\mathrm{O}},k\alpha)$ reaction was employed to search for the $k\alpha$-condensed states up to $k=6$,
but no experimental signatures were observed.
The authors of Ref.~\cite{Bishop2019} suggested that $k\alpha$-condensed states might be forbidden from decaying through the $k\alpha$ channels
due to their low decay energies and the Coulomb barriers.
The authors of Ref.~\cite{Barbui2018} attempted to determine the excitation function of $^{24}$Mg in the $^{20}\mathrm{Ne}+\alpha$ reaction on the basis of the thick target method.
They analyzed events where 4$\alpha$ particles were detected under the assumption that those $\alpha$ particles were emitted from $^{16}$O$^{*}$ excited
by the $^{20}\mathrm{Ne}+\alpha\to{^{16}\mathrm{O}^{*}}+{^{8}\mathrm{Be}(\mathrm{g.s.})}$ reaction.
They reported that a peak was observed at $E_x=34$~MeV in $^{24}$Mg close to the theoretically predicted energy of the 6$\alpha$-condensed state \cite{Yamada2004}
when they selectively analyzed events where the invariant mass of the detected 4$\alpha$ particles was close to that of the candidate for the 4$\alpha$-condensed state.
However, the peak significance remained highly uncertain due to the statistical limitation
and potential ambiguity in assumptions made in the analysis.

In the present work, we searched for the 6$\alpha$-condensed state in $^{24}$Mg by measuring the $\12C+\12C$ scattering.
Especially, we focused on events where 3$\alpha$ particles were emitted from the $0^+_2$ state in $\12C$
because we expected the 6$\alpha$-condensed state to decay through the 3$\alpha$-condensed state.
Previously, the $\12C+\12C\to\12C(0^+_2)+\12C(0^+_2)$ reaction was measured in Refs.~\cite{Wuosmaa1993,Wuosmaa1994},
and the candidate for the 6$\alpha$-chain state in $^{24}$Mg at $\Ecm=32.5$~MeV was proposed.
This 32.5-MeV resonance was investigated via various channels of the $\12C+\12C$ scattering,
and the strenuous attempts to clarify its internal structure were performed on the basis of the angular-momentum analysis \cite{Rae1992,Chappell1996,Marechal1997,Chappell1998}.
However, the excitation-energy range of $^{24}$Mg scanned for investigating the 32.5-MeV resonance was higher than that of the 6$\alpha$-condensed state.
In the lower excitation-energy domain,
the effective excitation function of the $\12C+\12C\to3\alpha+X$ reaction was reported from the indirect measurement of 3$\alpha$ particles in Ref.~\cite{Kolata1980}.
A remarkable peak structure was observed at $\Ecm\sim19.5$~MeV,
which corresponds to the theoretically predicted excitation energy of $E_x\sim33.4$~MeV in $^{24}$Mg.
However, it was not compelling evidence of the 6$\alpha$-condensed state because the 3$\alpha$ particles were not directly detected and their invariant mass was not determined.
In the recent measurement of the $\12C+\12C\to3\alpha+\12C$ reaction at $\Ecm=8.9$--21~MeV \cite{Wang2022},
the 3$\alpha$ particles emitted from the $\12C+\12C$ scattering were measured using the active-target time projection chamber.
It was reported that the excitation function of the $\12C+\12C\to3\alpha+X$ reaction tended to agree with the result in Ref.~\cite{Kolata1980} at energies above $\Ecm=19$~MeV.
However, the measurement was statistically too poor to observe the peak at $\Ecm\sim19.5$~MeV,
and the invariant mass of the 3$\alpha$ particles was not determined.
Thus, it was strongly desired to directly measure the 3$\alpha$ particles emitted from the $\12C+\12C$ scattering
and to determine their invariant mass for examining their decay histories.

The experiment was conducted in two beam-time periods at the R5 beam line of the Tokai Tandem accelerator facility of the Japan Atomic Energy Agency (JAEA).
We utilized a $\12C$ beam with an incident energy of $E_{\mathrm{lab}}=35.0$--50.0~MeV to explore the excitation-energy range of 31.4--38.9~MeV in $^{24}$Mg.
A $\12C$ beam bombarded the $^{\mathrm{nat}}$C target with a thickness of 100 $\mu$g/cm$^2$, and decay particles emitted from the $\12C+\12C$ collision were detected.
For the decay particle detector, we used the Si detector array SAKRA which was placed at the forward angle.
SAKRA was a lampshade shaped Si telescope array consisted of five Si-strip sensors, which were Design MMM fan-shaped Si sensors from Micron Semiconductor.
The thickness of the Si sensor was 500~$\mu$m. Its sector angle, outer radius and inner radius were 60 degrees, 135~mm, and 33~mm, respectively.
The junction side of this sensor was divided into 16 ring strips whereas the ohmic side was divided into 8 radial strips.
The configuration of SAKRA was optimized for pulse shape analysis (PSA)
by using neutron transmutation doped Si sensors with a special setup where charged particles were injected from the ohmic side \cite{LeNeindre2013}.
The charge signals from the Si sensors were processed with the Mesytec MPR-16/32 charge-sensitive preamplifiers
and digitized by the CAEN V1730SB 500~MS/s waveform digitizers.
At some beam energies,
several Si sensors of SAKRA did not properly work due to the shortage of bias voltage caused by the increase in the leak current.
Therefore, these sensors were excluded from the analysis at those energies.

\begin{figure}
  \centering
  \includegraphics[width=\Width]{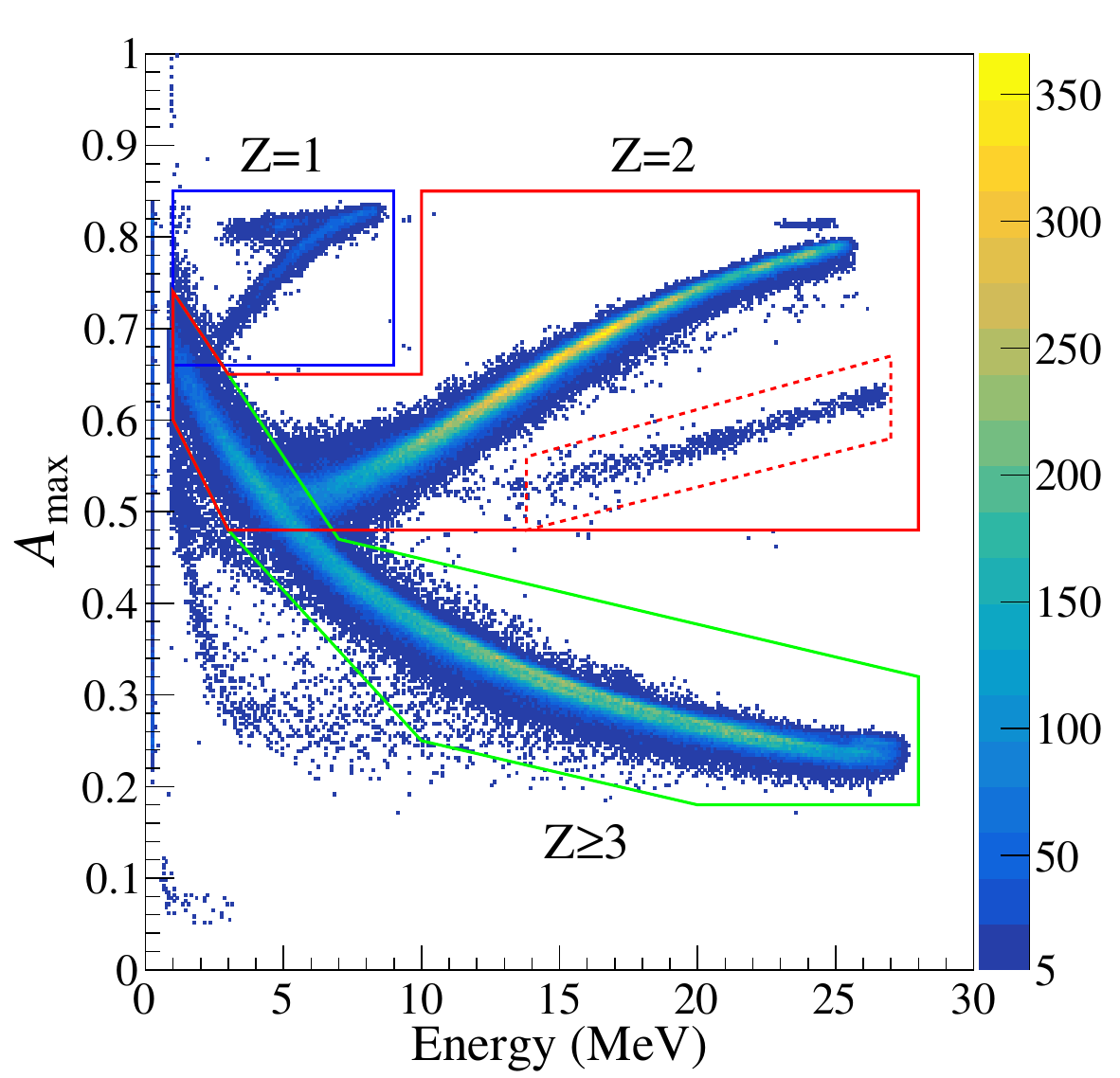}
  \caption{Correlation between $\Amax$ and kinetic energies of decay particles.
    The blue, red, and green boxes present the areas corresponding to charged particles with $Z=1$, $Z=2$, and $Z\geq3$, respectively.}
  \label{Fig:Amax}
\end{figure}

We performed particle identification (PID) for decay particles using the PSA.
We used the amplitude of the differential waveform of output signals from the preamplifier, named $\Amax$, as a PID parameter.
Figure~\ref{Fig:Amax} shows the correlation between $\Amax$ and kinetic energies of decay particles.
The loci in the blue, red, and green boxes correspond to charged particles with $Z=1$, $Z=2$, and $Z\geq3$, respectively.
The dominant isotopes in individual loci are proton, alpha, and $\12C$.
The locus in the red-dashed box corresponds to $^{8}$Be, namely, 2$\alpha$ particles detected in the same junction-side strip with almost same energy.
At lower kinetic energy than 5~MeV, alpha particles cannot be separated from $\12C$.
In this PID procedure with the PSA, only decay particles identified as non-alpha particles were excluded from the analysis
in order to prevent loss of low-energy alpha particles.

\begin{figure}
  \centering
  \includegraphics[width=\Width]{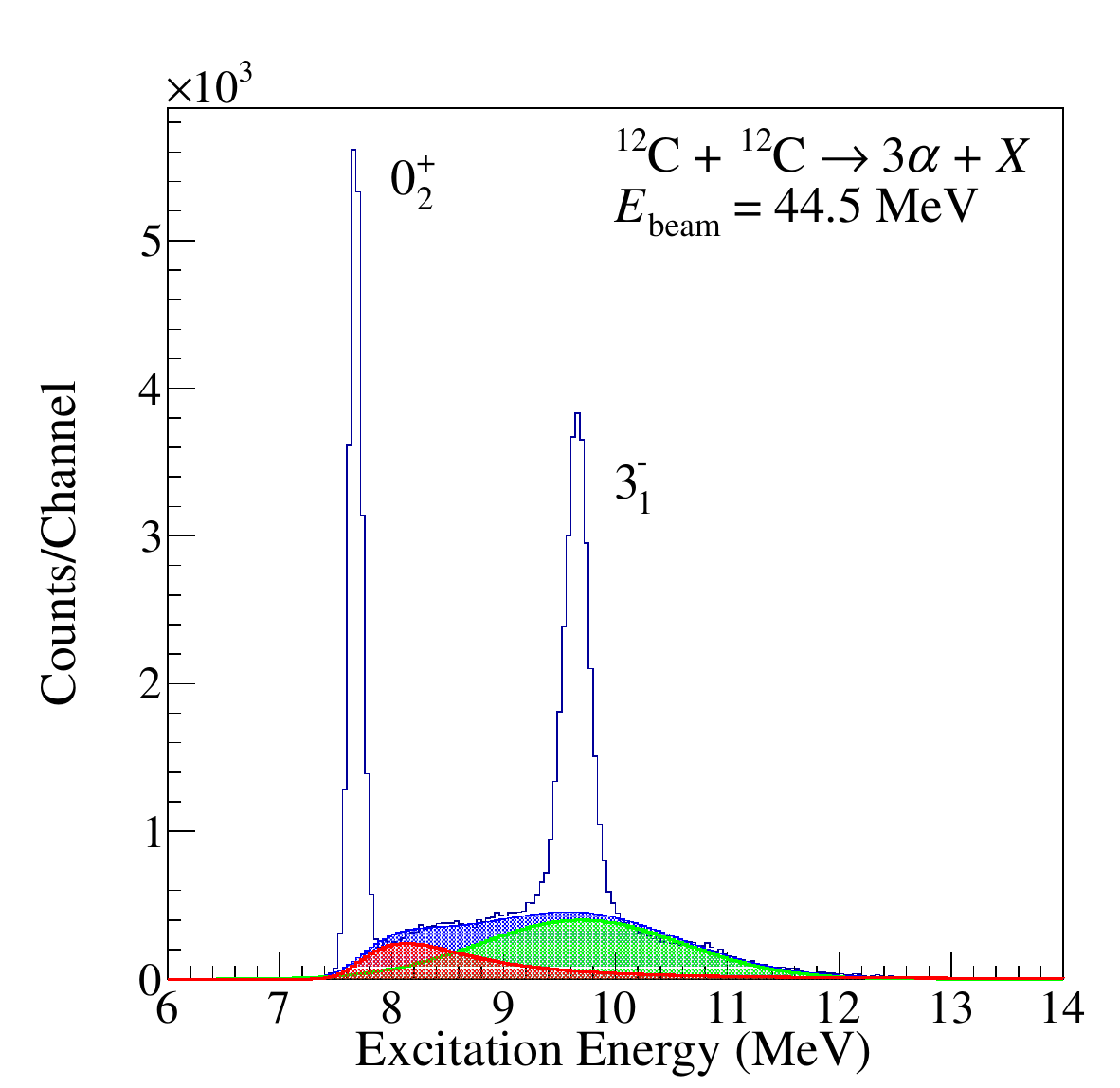}
  \caption{Excitation-energy spectrum of $\12C$ reconstructed from the detected 3$\alpha$ particles in the $\12C+\12C\to3\alpha+X$ reaction at $E_{\mathrm{beam}}=44.5$~MeV.
    The green spectrum shows the Gaussian function representing the broad states in $\12C$
    whereas the red spectrum is the Landau distribution function representing the other backgrounds.    
    The sum of two spectra is shown by the blue spectrum.
  }
  \label{Fig:inv}
\end{figure}

In the events involving 3$\alpha$ particles emitted from the same $\12C$ nucleus,
these particles were boosted along the velocity of the center of mass, and likely to hit a single Si sensor.
In order to identify the decay events via the 3$\alpha$-condensed state,
we reconstructed the invariant mass of the 3$\alpha$ particles detected in a single Si sensor,
and determined the excitation energy of parent $\12C$.
Figure~\ref{Fig:inv} shows the excitation-energy spectrum of $\12C$ reconstructed from the detected 3$\alpha$ particles in the $\12C+\12C\to3\alpha+X$ reaction
at $E_{\mathrm{beam}}=44.5$~MeV.
In Fig.~\ref{Fig:inv}, the two prominent peaks due to the $0^+_2$ state at 7.65~MeV and the $3^-_1$ state at 9.64~MeV are observed
on the continuous spectrum due to the broad $0^+_3$ and $2^+_2$ state in $\12C$ \cite{Itoh2011} and the other backgrounds.
The backgrounds were mainly originated from events in which the PID was incorrect,
while contribution from accidental coincidence events was negligible.
The continuous spectrum was fitted by the combination of a Gaussian function representing the broad $\12C$ states
and a Landau distribution function representing the backgrounds.
The green and red spectra show the Gaussian and Landau distribution functions, respectively,
whereas the blue spectrum is the sum of them.
The parameters of the two functions were determined to reproduce the experimental spectrum.
The yields for the $\12C+\12C\to\12C(0^+_2)+X$ and $\12C+\12C\to\12C(3^-_1)+X$ reactions were estimated
by subtracting the blue spectrum from the measured spectrum.

\begin{figure}
  \centering
  \includegraphics[width=\Width]{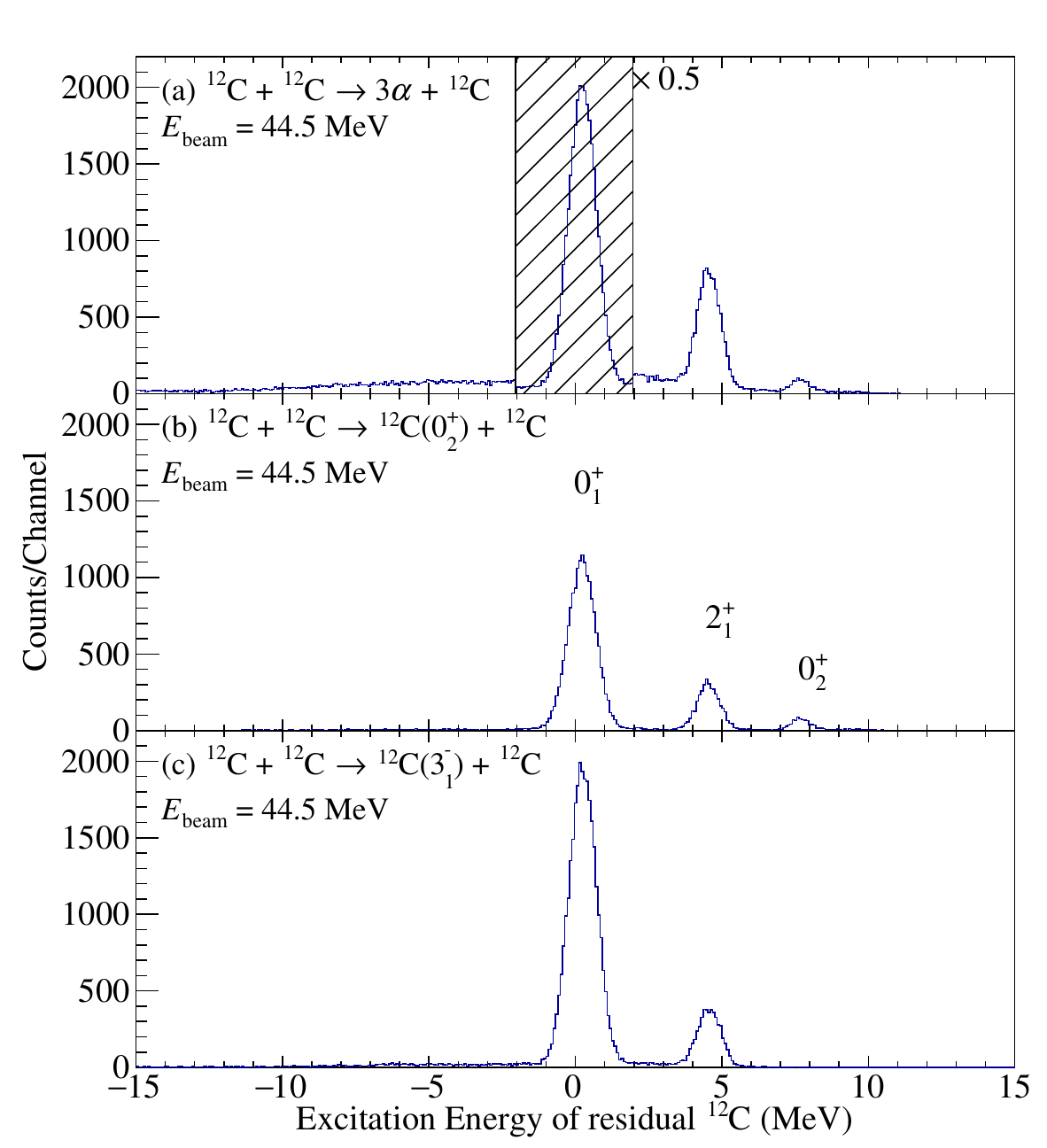}
  \caption{Excitation-energy spectra of the residual $\12C$ for (a) $\12C+\12C\to3\alpha+\12C$,
    (b) $\12C+\12C\to\12C(0^+_2)+\12C$, and (c) $\12C+\12C\to\12C(3^-_1)+\12C$ reactions.
    The spectrum in the hatched area is downscaled by a factor of 0.5.
  }
  \label{Fig:mis}
\end{figure}

In addition, in order to exclusively identify reaction channels,
we determined the excitation energy of the residual $\12C$ nucleus by estimating the missing mass of the reaction from momenta of the detected 3$\alpha$ particles.
Figure~\ref{Fig:mis}(a) shows the excitation-energy spectra of the residual $\12C$ for all the events presented in Fig.~\ref{Fig:inv},
whereas Figs.~\ref{Fig:mis}(b) and (c) are the excitation-energy spectra for the selected events
in which the excitation energy in Fig.~\ref{Fig:inv} is close to those of the $0^+_2$ and $3^-_1$ states, respectively.
Although a little continuous background was observed below the prominent 3 peaks
corresponding to the $0^+_1$, $2^+_1$, and $0^+_2$ states of the residual $\12C$ nuclei in Fig.~\ref{Fig:mis}(a),
there was almost no continuum background in Figs.~\ref{Fig:mis}(b) and (c) after the invariant masses of the detected 3$\alpha$ particles were selected.
The yields for the $\12C+\12C\to\12C(0^+_2)+\12C(0^+_1)$, $\12C(0^+_2)+\12C(2^+_1)$, and $\12C(0^+_2)+\12C(0^+_2)$ reactions were determined
from Fig.~\ref{Fig:mis}(b).

To estimate the detection efficiency for each reaction channel,
the Monte Carlo simulations were performed in the following three steps.
In all the steps of the simulation, we assumed that two $\12C$ nuclei were emitted from the $\12C+\12C$ reaction.
If the $\12C$ nucleus was excited to the $0^+_2$ or $3^-_1$ states,
it would sequentially decayed to 3$\alpha$ particles via the $\alpha+{^{8}\mathrm{Be(g.s.)}}$ channel.
The two-body decay in each step isotropically occurred in the rest frame of the parent nucleus.
We analyzed the simulated data in the same manner as the experimental data,
and estimated the efficiency of 3$\alpha$ detection and $\12C$ reconstruction for each reaction channel.

In the first step,
all the solid angles were divided  into polar angular bins with a width of $\theta_{\mathrm{cm}}=4.0^{\circ}$,
and the efficiency for individual reaction channels at every beam energy was estimated
in each angular bin in order to determine the differential cross section.
We assumed that the differential cross section was constant within one angular bin.
In each reaction channel, experimental data at beam energies where the maximum value of the efficiency at each angular bin was below 5\% were excluded from the analysis.

In the second step, the averaged efficiency over the angular acceptance $\Delta\Omega_{\mathrm{cm}}$ for individual reaction channels was estimated at each beam energy.
In general, the angular acceptance in the center-of-mass frame declines as the beam energy decreases
because the decay particles are emitted over a wider angular range in the laboratory frame at the lower center-of-mass energy.
Therefore, the angular acceptance of each reaction channel was determined at the lowest beam energy to include only those angular bins
where their efficiency exceeded 1\%.
For example, the angular acceptance for the $\12C+\12C\to\12C(0^+_2)+\12C(0^+_1)$, $\12C(0^+_2)+\12C(2^+_1)$, and $\12C(0^+_2)+\12C(0^+_2)$ reactions
were $44^{\circ}\leq\theta_{\mathrm{cm}}\leq 144^{\circ}$, $52^{\circ}\leq\theta_{\mathrm{cm}}\leq 144^{\circ}$, and $68^{\circ}\leq\theta_{\mathrm{cm}}\leq 140^{\circ}$, respectively.
The second simulation was conducted using the measured angular distributions of the differential cross sections for the individual reaction channels obtained in the first step.
The cross sections integrated over the angular acceptances were obtained by using the total yields and the angular averaged efficiency.
For comparing the cross sections of various reaction channels with the different angular acceptances,
the obtained cross sections were normalized by multiplying a factor of $4\pi/\Delta\Omega_{\mathrm{cm}}$.

In the third step, the averaged efficiency over all the solid angles for the inclusive $\12C+\12C\to\12C(0^+_2\mathrm{~or~}3^-_1)+X\to3\alpha+X$ reactions was estimated.
In this simulation, the $0^+_1$, $2^+_1$, and $0^+_2$ states were taken into account as the final states of the residual $\12C$ nuclei.
The population of the residual states was determined from the exclusive cross sections for the various reaction channels obtained in the second step.
However, unlike in the second step, the angular distributions of the differential exclusive cross sections were assumed to be isotropic
because the angular acceptances $\Delta\Omega_{\mathrm{cm}}$ applied in the second step were different between the reaction channels.
Finally, the total cross sections of the inclusive reactions were obtained.

\begin{figure}
  \centering
  \includegraphics[width=\Width]{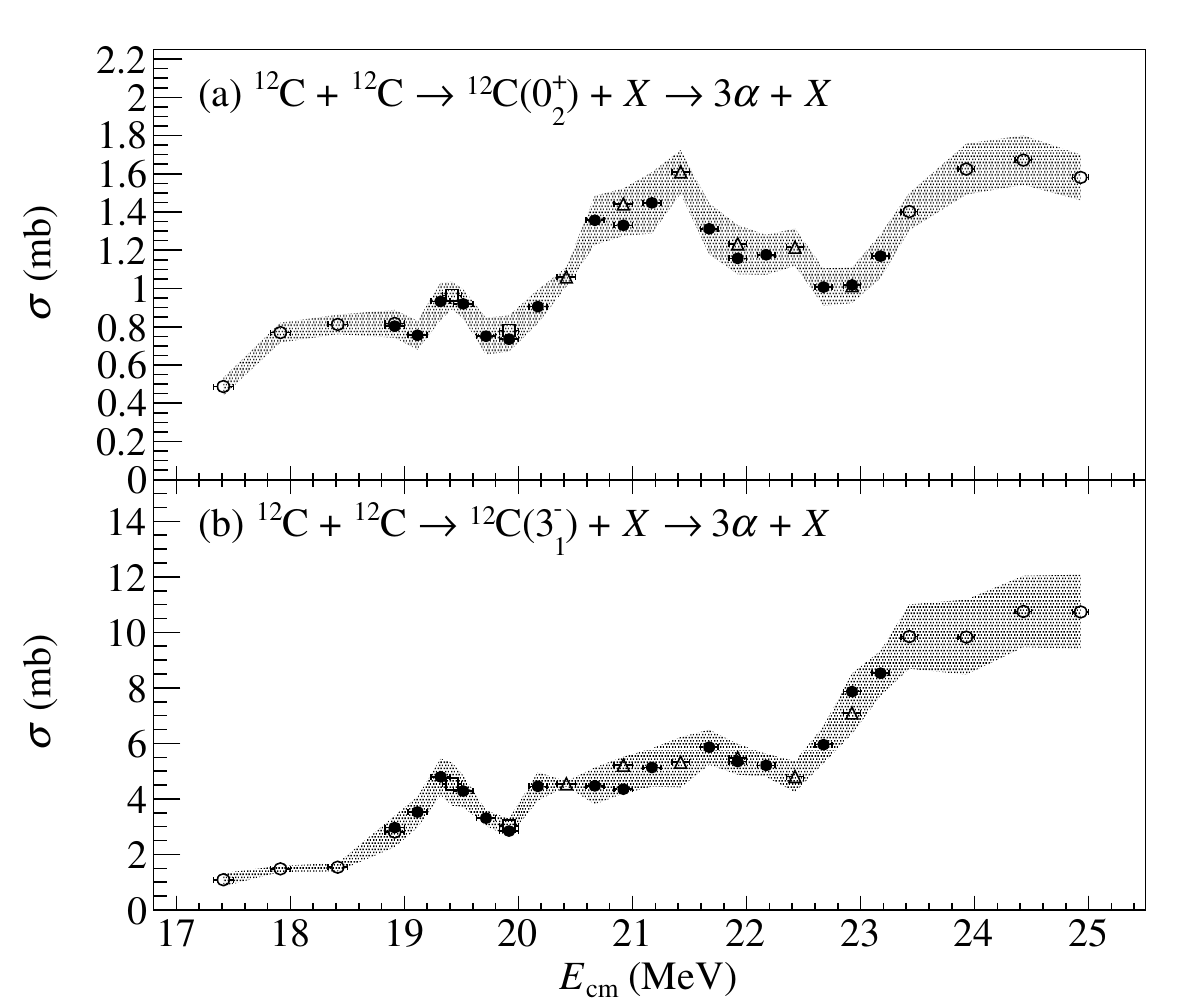}
  \caption{Inclusive cross sections for the (a) $\12C+\12C\to\12C(0^+_2)+X\to3\alpha+X$ and (b) $\12C+\12C\to\12C(3^-_1)+X\to3\alpha+X$ reactions.
    The circles, squares, and triangles denote the experimental data acquired with the 5, 4, and 3 Si sensors, respectively.
    The open symbols correspond to the results from the first beam-time period,
    and the solid symbols represent those from the second beam-time period.
  }
  \label{Fig:incl}
\end{figure}

The inclusive cross sections for the $\12C+\12C\to\12C(0^+_2\mathrm{~or~}3^-_1)+X\to3\alpha+X$ reactions
are plotted as a function of the center-of-mass energy $\Ecm$ of the initial $\12C+\12C$ system in Fig.~\ref{Fig:incl}.
The cross sections were acquired by using reliable segments of SAKRA at each beam energy.
The circle, rectangle, and triangle symbols denote that the number of the used segments was 5, 4, and 3, respectively.
The open symbols correspond to the results from the first beam-time period, and the solid symbols represent those from the second beam-time period.
It was found that the absolute values of the cross sections systematically fluctuated among the five segments of SAKRA
although the energy dependence of the cross section obtained by each single segment was similar.
Therefore, we treated these fluctuations as the systematic uncertainties and show them as the gray bands in Fig.~\ref{Fig:incl}.

A relatively narrow peak structure is observed at $\Ecm=19.4$~MeV in both Figs.~\ref{Fig:incl}(a) and (b),
while the broad structure is observed at $\Ecm=20.0$--22.0~MeV in Fig.~\ref{Fig:incl}(a) only.
It should be noted that the narrower state at $\Ecm=19.4$~MeV had been previously observed in indirect measurements of the 3$\alpha$ emission
from the $\12C+\12C$ reaction \cite{Kolata1980}.
The present work showed that this state decayed to the 3$\alpha$ particles via the $0^+_2$ and $3^-_1$ states in $\12C$ by the direct measurement.
This is strong evidence that this narrower state is a multi-$\alpha$ cluster state
because the $0^+_2$ and $3^-_1$ states are known as the spatially developed 3$\alpha$-cluster states.
This 19.4-MeV state actually couples with the $0^+_2$ state as the 3$\alpha$-condensed state,
and its energy of $\Ecm=19.4$~MeV, corresponding to the excitation energy of $E_x=33.3$~MeV in $^{24}$Mg,
is close to the theoretical value of the 6$\alpha$-condensed state predicted in Ref.~\cite{Yamada2004}.
It is, therefore, reasonable to consider this state as a candidate for the 6$\alpha$-condensed state.
However, the spin and parity of this candidate remain unknown.
It is highly desirable to determine the spin and parity of this state.

\begin{figure}
  \centering
  \includegraphics[width=\Width]{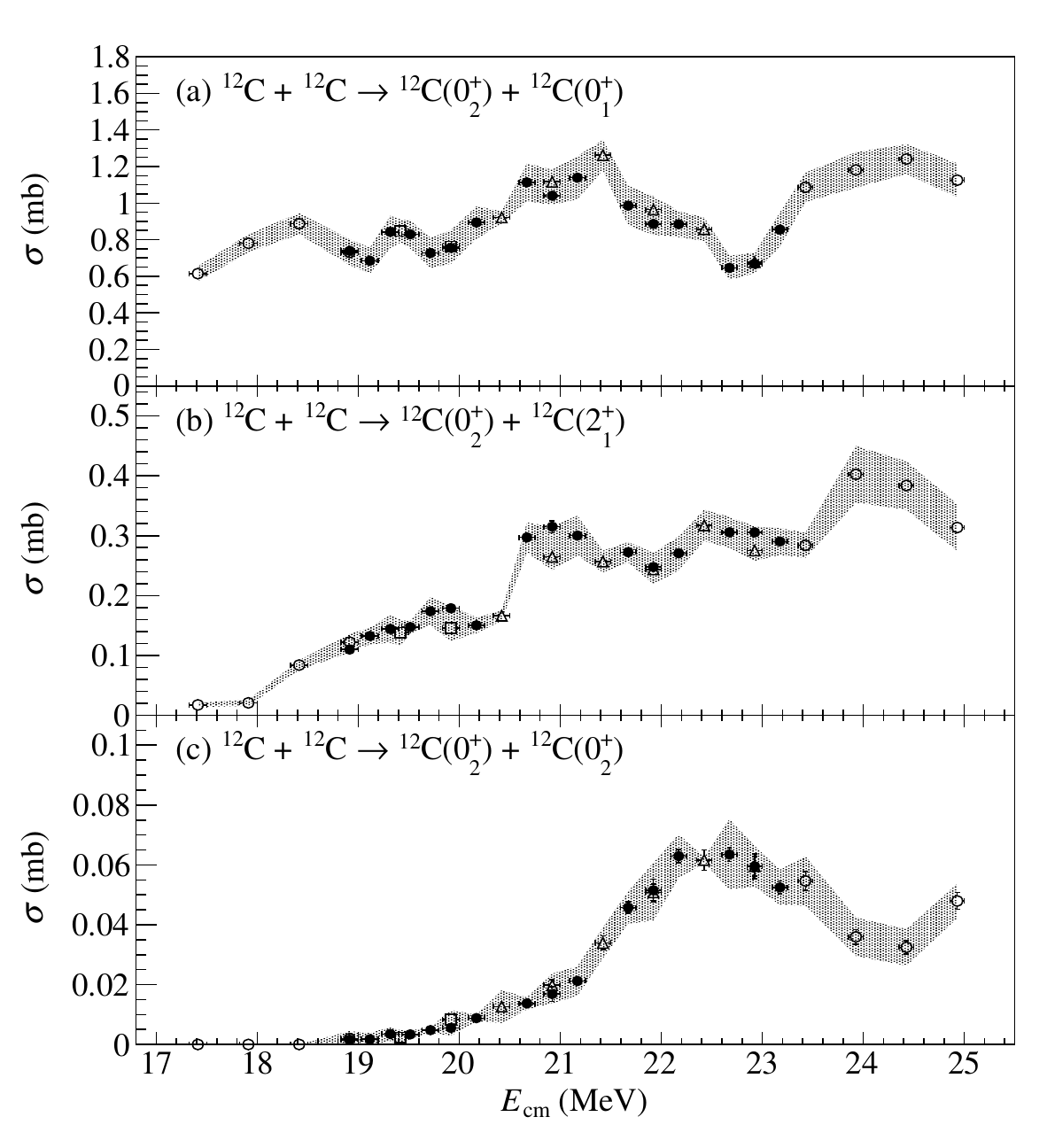}
  \caption{
    Exclusive cross sections for the $\12C+\12C\to$ (a) $\12C(0^+_2)+\12C(0^+_1)$, (b) $\12C(0^+_2)+\12C(2^+_1)$, and (c) $\12C(0^+_2)+\12C(0^+_2)$ reactions.
    The notation of the symbols were same as Fig.~\ref{Fig:incl}.
  }
  \label{Fig:excl}
\end{figure}

Figures~\ref{Fig:excl}(a), (b), and (c) show the exclusive cross sections
for the $\12C+\12C\to\12C(0^+_2)+\12C(0^+_1)$, $\12C(0^+_2)+\12C(2^+_1)$, and $\12C(0^+_2)+\12C(0^+_2)$ reactions, respectively.
The energy dependence of the measured cross sections exhibits different behavior for each reaction channel.
This indicates the successful separation of the reaction channels in the present analysis.
The peak at $\Ecm=19.4$~MeV, which is a candidate for the 6$\alpha$-condensed state,
is observed in the $\12C(0^+_2)+\12C(0^+_1)$ channel in Fig.~\ref{Fig:excl}(a),
but is not in the $\12C(0^+_2)+\12C(0^+_2)$ channel in Fig.~\ref{Fig:excl}(c).
This result appears to contradict the expectation that the overlap of the wave functions between the 6$\alpha$-condensed state and $\12C(0^+_2)+\12C(0^+_2)$ should be larger rather than
that between the 6$\alpha$-condensed state and $\12C(0^+_2)+\12C(0^+_1)$.
A plausible explanation of this result is that the $\12C(0^+_2)+\12C(0^+_2)$ channel is hindered
by the Coulomb barrier due to the low decay energy as pointed out in Ref.~\cite{Bishop2019}.
A broad peak is observed around $\Ecm=22.5$~MeV in the cross-section spectrum for the $\12C+\12C\to\12C(0^+_2)+\12C(0^+_2)$ reaction shown in Fig.~\ref{Fig:excl}(c).
This energy corresponds to the excitation energy of 36.4~MeV in $^{24}$Mg.
This broad state ought to be a multi-$\alpha$ cluster state as well as the narrow state found at $\Ecm=19.4$~MeV.
Because the $\12C(0^+_2)+\12C(0^+_2)$ channel should be strongly coupled with the 6$\alpha$-condensed state,
this broad state should be akin to the 6$\alpha$-condensed state.
This state is inferred to be an excited state of the 6$\alpha$-condensed state such as the $2^+_2$ and $4^+_1$ states in $\12C$ \cite{Funaki2015}
since its excitation energy is 3.1~MeV higher than the theoretical value of the 6$\alpha$-condensed state predicted in Ref.~\cite{Yamada2004}.
In the $\12C$ case,
the excited states of the 3$\alpha$-condensed state with $J^\pi=2^+$ and $4^+$ were theoretically predicted at 1.4 and 3.5~MeV
above the 3$\alpha$-condensed state \cite{Funaki2015},
and their experimental counterparts were reported in Refs.~\cite{Itoh2011,Zimmerman2013,Freer2011}.

\begin{figure}
  \centering
  \includegraphics[width=\Width]{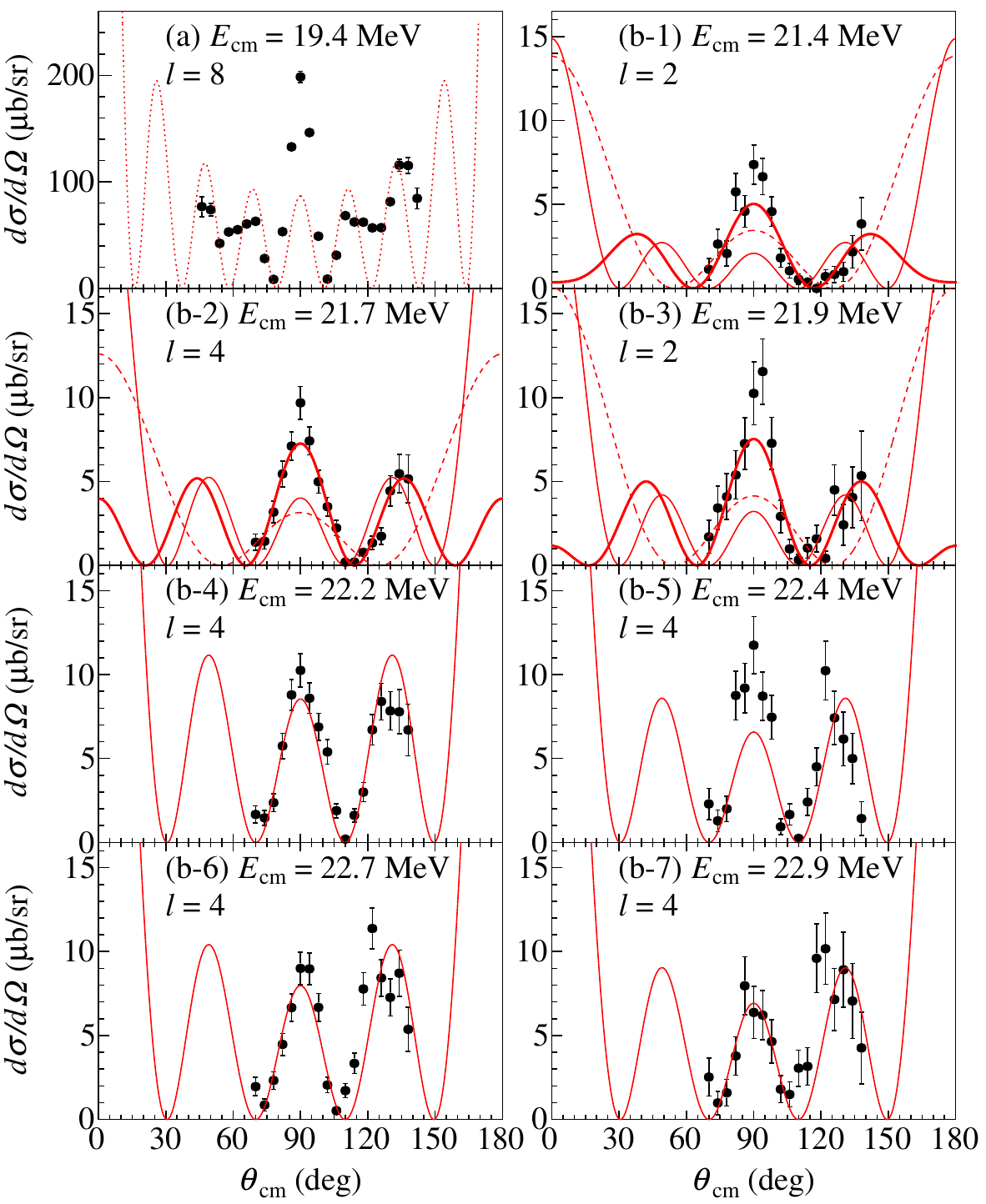}
  \caption{
    Angular distributions of the differential cross sections for the $\12C+\12C\to$ (a) $\12C(0^+_2)+\12C(0^+_1)$ and (b) $\12C(0^+_2)+\12C(0^+_2)$ reactions.
    The dashed, thin solid, and dotted lines show squared Legendre polynomials of order 2, 4, and 8, respectively.
    The best-fit $l$ value is indicated in each panel.
    In (b-1)--(b-3),
    the Legendre polynomial fits taking contributions from $l=$ 2 and 4 into account are shown as the thick solid lines.
  }
  \label{Fig:angle_dist}
\end{figure}

To investigate the spins and parities of the observed states,
we analyzed the angular distributions of the differential cross sections by fitting them by a squared Legendre polynomial of order $l$.
Figure~\ref{Fig:angle_dist}(a) shows the angular distribution of the differential cross sections for the $\12C+\12C\to$ $\12C(0^+_2)+\12C(0^+_1)$ reaction at $\Ecm=19.4$~MeV.
Its diffraction pattern is best fitted by the Legendre polynomial of order 8 shown by the dotted line
although the enhancement of the diffraction amplitude at $\theta_\mathrm{cm}=90^\circ$ is not reproduced.
This result means that the cross section around $\Ecm=19.4$~MeV is dominated by the $L=8$ component
which is very close to the grazing angular moment of $L_g=8.8$,
and it does not suggest the $L=0$ component populating the 6$\alpha$-condensed state with $J^\pi=0^+$.
This is due to the fact that contamination from continuous components was quite large as seen in Fig~\ref{Fig:excl}(a),
and it obscures the $L=0$ component even if the 19.4-MeV state is the 6$\alpha$-condensed state.
The enhancement of the differential cross section around $\theta_\mathrm{cm}=90^\circ$ also indicates
that the cross section around $\Ecm=19.4$~MeV is dominated by the continuous components with several multipolarities components
because this enhancement is not reproduced with a single Legendre polynomial but only by introducing several partial waves.
Therefore,
the subtraction of the continuous components is necessary to determine the spin and parity of the small peak on the continuous spectrum.
However, its yield was statistically too poor to reliably perform this analysis, and we could not determine the spin and parity of this state.

On the other hand, as shown in Figs.~\ref{Fig:angle_dist}(b-4)--(b-7),
the angular distributions for the $\12C+\12C\to$ $\12C(0^+_2)+\12C(0^+_2)$ reaction at four energies around $\Ecm=22.5$~MeV were well represented
with a single Legendre polynomial of order 4.
It is reasonable to consider that the spin and parity of the broad state around $\Ecm=22.5$~MeV is $4^+$.
At lower energies,
it is difficult to reproduce the angular distributions with a single Legendre polynomial due to the enhancement around $\theta_\mathrm{cm}=90^\circ$
seen in Figs.~\ref{Fig:angle_dist}(b-1)--(b-3).
The angular distributions at $\Ecm=21.4$, 21.7, and 21.9~MeV were fitted with the formula
$d\sigma/d\Omega(\theta_\mathrm{cm})=|\sum_{l=2,4} a_l e^{i\phi(l)}P_l(\cos\theta_\mathrm{cm})|^2$ taken from Ref.~\cite{Wuosmaa1994}.
In Figs.~\ref{Fig:angle_dist}(b-1)--(b-3),
the best-fit results with this formula are shown by the thick solid lines,
those with a single Legendre polynomial of order 2 or 4 are plotted by the dashed or thin solid lines, respectively.
The angular distributions are better fitted by incorporating both the $L=2$ and $L=4$ components into account
than by utilizing a single Legendre polynomial.
It is a signature that a $2^+$ state exists at the low-energy side of the broad state around $\Ecm=22.5$~MeV
although the $2^+$ resonance could not be separated from the broad $4^+$ state in the energy spectrum in Fig.~\ref{Fig:excl}(c).
These spin-parity assignments are consistent with our speculation that this broad structure is due to excited states of the 6$\alpha$-condensed state.
Considering the difference of the excitation energy and their spins and parities,
these states might be excited states of the 6$\alpha$-condensed state in similar to the $2^+_2$ and $4^+_1$ states in $\12C$.

In summary, we measured the $\12C+\12C$ scattering using the SAKRA detector in order to search for the 6$\alpha$-condensed state in $^{24}$Mg.
The inclusive cross sections for the $\12C+\12C\to\12C(0^+_2\mathrm{~or~}3^-_1)+X\to3\alpha+X$ reactions peaked at $\Ecm=19.4$~MeV,
which corresponds to the excitation energy of 33.3~MeV in $^{24}$Mg.
This 19.4-MeV state was observed in the direct measurement of the 3$\alpha$ emission from the $\12C+\12C$ reaction for the first time,
and was found to decay via the $0^+_2$ and $3^-_1$ states in $\12C$.
This fact is strong evidence that the 19.4-MeV state is a multi-$\alpha$ cluster state.
This 19.4-MeV state actually couples with the $0^+_2$ state as the 3$\alpha$-condensed state, 
and its energy is close to the theoretical value of the 6$\alpha$-condensed state.
These results suggest that this state is a candidate for the 6$\alpha$-condensed state.
This peak was also observed in the exclusive $\12C+\12C\to\12C(0^+_2)+\12C(0^+_1)$ reaction,
but not in the $\12C+\12C\to\12C(0^+_2)+\12C(0^+_2)$ reaction
although the 6$\alpha$-condensed state should have a larger overlap with the $\12C(0^+_2)+\12C(0^+_2)$ channel than with the $\12C(0^+_2)+\12C(0^+_1)$ channel.
This could be explained by an inference that the $\12C(0^+_2)+\12C(0^+_2)$ channel is suppressed by the Coulomb barrier due to its low decay energy.
In the $\12C(0^+_2)+\12C(0^+_2)$ channel, we found a broad structure around $\Ecm=22.5$~MeV, corresponding to $E_x=36.4$~MeV in $^{24}$Mg.
This broad state ought to be a multi-$\alpha$ cluster state as well as the narrow state found at $\Ecm=19.4$~MeV.

To provide further clarification regarding the internal structure of these observed states,
the spins and parities of these states are required.
The angular distributions of the cross sections were analyzed to gain an insight into the spin and parity.
Although the spin and parity of the 19.4-MeV state could not be resolved due to the continuous components,
those of the broad 22.5-MeV state were assigned to be $4^+$.
In addition, a signature of the $2^+$ state was suggested at lower-energy side of the broad state.
Considering the excitation energy which is 3.1~MeV higher than the theoretical value of the 6$\alpha$-condensed state,
this broad state might be akin to the $\alpha$-condensed state like the Hoyle band states known as the $2^+_2$ and $4^+_1$ states in $\12C$.

It is also essential to compare the present results with theoretical calculations.
However no microscopic calculations for the 6$\alpha$-condensed state are available at present.
Therefore, theoretical efforts are strongly desired to establish the 6$\alpha$-condensed state.

The authors acknowledge the JAEA tandem accelerator crews for providing a high-quality beam.
Y.~F. appreciates the support from the JSPS Research Fellowships for Young Scientists.
This work was partly supported by JSPS KAKENHI Grants No. JP19J20784, JP19H05604, JP19H05153, and JP21H00123.

\bibliographystyle{elsarticle-num.bst}

\bibliography{../mybib}

\end{document}